\documentclass[10pt,conference]{IEEEtran}
\IEEEoverridecommandlockouts
\usepackage{cite}
\usepackage{amsmath,amssymb,amsfonts}
\usepackage{graphicx}
\usepackage{textcomp}
\usepackage{xcolor}
\usepackage{algpseudocode}
\usepackage{algorithm,algorithmicx,algcompatible}
\usepackage{braket}
\usepackage{booktabs,multirow}

\def\BibTeX{{\rm B\kern-.05em{\sc i\kern-.025em b}\kern-.08em
    T\kern-.1667em\lower.7ex\hbox{E}\kern-.125emX}}

\usepackage{adjustbox}
\usepackage{tikz}
\usetikzlibrary{quantikz}
\usetikzlibrary{external}
\newcommand{\dicke}[2]{\ket{\smash{D_{#2}^{#1}}}}
\newcommand{\dsu}[1]{\textit{DSU}(#1)}
\newcommand{\ryangle}[2]{\theta_{\scriptstyle\surd#1/#2}}
\newcommand{\rygate}[2]{\gate{R_y(\ryangle{#1}{#2})}}

\usepackage[colorlinks=true]{hyperref}

\usepackage[switch]{lineno}
\newcommand*\patchAMSlineno[1]{
	\expandafter\let\csname old#1\expandafter\endcsname\csname #1\endcsname
	\expandafter\let\csname oldend#1\expandafter\endcsname\csname end#1\endcsname
	\renewenvironment{#1}
		{\linenomath\csname old#1\endcsname}
		{\csname oldend#1\endcsname\endlinenomath}
}
\AtBeginDocument{
	\patchAMSlineno{equation}
	\patchAMSlineno{equation*}
	\patchAMSlineno{align}
	\patchAMSlineno{align*}
}

\begin{document}


\title{Quantum Telecloning on NISQ Computers\\
{}
\thanks{Research presented in this article was supported by the Laboratory Directed Research  and  Development  program  of  Los  Alamos  National  Laboratory under project number 20220656ER.\hfill LA-UR-22-23967}
}

\author{
    \IEEEauthorblockN{Elijah Pelofske\IEEEauthorrefmark{1}\IEEEauthorrefmark{2}, 
    Andreas Bärtschi\IEEEauthorrefmark{1}\IEEEauthorrefmark{2}, 
    Bryan Garcia\IEEEauthorrefmark{3},
    Boris Kiefer\IEEEauthorrefmark{3},
    Stephan Eidenbenz\IEEEauthorrefmark{2}}
    \IEEEauthorblockA{\IEEEauthorrefmark{2}
    \textit{CCS-3 Information Sciences, Los Alamos National Laboratory},
	Los Alamos, NM 87544, USA
	}
    \IEEEauthorblockA{\IEEEauthorrefmark{3}
    \textit{Department of Physics, New Mexico State University}, Las Cruces, NM 88003, USA
    }
    \IEEEauthorblockA{\IEEEauthorrefmark{1}
    Corresponding authors:
    \href{mailto:epelofske@lanl.gov}{epelofske@lanl.gov}, 
	\href{mailto:baertschi@lanl.gov}{baertschi@lanl.gov}
	}
}

\maketitle

\begin{abstract}
	Due to the no-cloning theorem, generating perfect quantum clones of an arbitrary unknown quantum state is not possible, however approximate quantum clones can be constructed. Quantum telecloning is a protocol that originates from a combination of quantum teleportation and quantum cloning. Here we present $1 \rightarrow 2$ and $1 \rightarrow 3$ quantum telecloning circuits, with and without ancilla, that are theoretically optimal (meaning the clones have the highest fidelity allowed by quantum mechanics), universal (meaning the clone fidelity is independent of the state being cloned), and symmetric (meaning the clones all have the same fidelity). We implement these circuits on gate model IBMQ and Quantinuum NISQ hardware and quantify the clone fidelities using parallel single qubit state tomography. Quantum telecloning using mid-circuit measurement with classical feed-forward control (i.e. real time if statements) is demonstrated on the Quantinuum H1-2 device. Two alternative implementations of quantum telecloning, deferred measurement and post selection, are demonstrated on ibmq\_montreal, where mid-circuit measurements with real time if statements are not available. Our results show that NISQ devices can achieve near-optimal quantum telecloning fidelity; for example the Quantinuum H1-2 device running the telecloning circuits without ancilla achieved a mean clone fidelity of $0.824$ with standard deviation of $0.024$ for two clone circuits and $0.765$ with standard deviation of $0.022$ for three clone circuits. The theoretical fidelity limits are $0.8\overline{3}$ for two clones and $0.\overline{7}$ for three clones. This demonstrates the viability of performing experimental analysis of quantum information networks and quantum cryptography protocols on NISQ computers.
\end{abstract}

\begin{IEEEkeywords}
Quantum computing, quantum telecloning, telecloning, NISQ computers, parallel state tomography, quantum cloning, measurement error mitigation, mid-circuit measurement, Qiskit, bell state measurement
\end{IEEEkeywords}

\section{Introduction}
\label{sec:introduction}

The no-cloning theorem \cite{wootters1982single} states that due to the linearity of quantum mechanics an arbitrary quantum state cannot be perfectly cloned, with the corollary that the laws of quantum mechanics preclude eavesdropping on quantum communication \cite{bennett1984proceedings}. Interestingly the no-cloning theorem does not preclude approximate cloning of a quantum state \cite{duan1998probabilistic, pati1999quantum, buvzek1996quantum}. 

Quantum cloning of $N$ initial copies of a 1-qubit quantum state into $M>N$ clones can be distinguished using several criteria. First, quantum cloning can be either \emph{universal} or \emph{state-dependent} \cite{fan2014quantum,gisin1997optimal,werner1998optimal,buvzek1998universal,bruss1998optimal}. Universal indicates that the cloning process will generate clones with equal fidelity, regardless of the input state that is being cloned \cite{bruss1998optimal,buvzek1998universal}. State dependent means that the fidelity of the clones that are generated is dependent on the input state being cloned. The second criteria is whether the cloning process is \emph{symmetric} or \emph{asymmetric}. Symmetric means that all clones have the same fidelity \cite{Braunstein_2001}, whereas asymmetric means that the clones can have different fidelities \cite{Niu_1998, Iblisdir_2005, PhysRevLett.84.4497, Zhao_2005, 10.5555/2011742.2011744}. The third criteria is whether the cloning process is \emph{optimal} \cite{buvzek1996quantum,Zhao_2005,Zhen_2008} or not. Here optimal means that the clones that are generated are of the highest fidelity allowed by quantum mechanics. The final criteria is whether the cloning is \emph{phase-covariant} \cite{fan2001quantum, zhang2020optimal}, which indicates that the input states lie on the equator of the Bloch sphere. For optimal universal symmetric cloning, the optimal fidelity bound of $M$ clones of $N$ initial states can be rigorously computed \cite{scarani2005quantum, PhysRevA.59.156}:
\begin{equation}
    F_{N \rightarrow M} = \frac{ MN + M + N }{ M(N+2) }
    \label{eq:theoretical-fidelity}
\end{equation}

In conjunction with quantum teleportation \cite{bennett1993teleporting, pirandola2015advances}, optimal cloning can be employed to transmit a quantum message $N$ to $M$ recipients in a process known as \emph{quantum telecloning} \cite{PhysRevA.61.032311, PhysRevA.67.012323}. This is done through Bell measurements of the message and port qubits followed by local Pauli X/Z operations on the recipients' clones based on classical communication. This telecloning protocol can be implemented on universal quantum computers \cite{forcer2002superposition, https://doi.org/10.48550/arxiv.quant-ph/9707034} if the classical communication can be emulated or implemented directly. In this article we consider telecloning of a single input state ($N=1$). The optimal fidelities for $1 \rightarrow 2$ and $1 \rightarrow 3$ telecloning are $\frac{5}{6}  \hspace{2mm} (0.8\overline{3})$ and $\frac{7}{9} \hspace{2mm} (0.\overline{7})$, respectively. In the remainder of this paper, all \emph{quantum telecloning} circuits are defined to specifically be \emph{optimal}, \emph{symmetric}, \emph{universal}, and not phase-covariant. 

Quantum cloning is relevant for a variety of reasons including as a key component of quantum network protocols \cite{PhysRevA.56.3446}. Quantum telecloning has been experimentally demonstrated in a variety of experimental settings \cite{Qiong_2008}, including as a quantum networking protocol \cite{PhysRevLett.109.173604, li2016photonic}, using trapped atom devices \cite{Zhen_2008}, for qudits (d-dimensional quantum systems) instead of qubits \cite{Araneda2016, Xin_Wen_2010, PhysRevLett.87.247901, zhang2020optimal}, and continuous variable (see \cite{Lloyd_1999}) implementations \cite{PhysRevA.104.032419, PhysRevLett.85.1754, PhysRevA.77.022316} and experimental proposals \cite{Gordon2007}. Remote information concentration, which is the inverse process of telecloning where distributed quantum information is concentrated into a single qubit, has also been investigated \cite{wang2013manytoone, PhysRevLett.86.352}. For a review on telecloning see \cite{fan2014quantum}, and for a review on quantum cloning see \cite{scarani2005quantum}.

\begin{figure*}[t!]
    \centering\footnotesize
    \begin{tabular}{@{}l rr rr r@{}r@{}}
        \toprule
        & \multicolumn{2}{c}{{\bf ibmq\_montreal} (deferred measurement)}
        & \multicolumn{2}{c}{{\bf ibmq\_montreal} (post selection)}
        & & {\bf Quantinuum H1-2} (mid-circuit measurement)
        \\
        \cmidrule(lr){2-3}
        \cmidrule(lr){4-5}
        \cmidrule{7-7}
        Telecloning
        & mean fidelity (SD) & meas.error mitigated
        & mean fidelity (SD) & m.err. mitigated
        & & mean fidelity (standard deviation SD)
        \\
		\midrule[\heavyrulewidth]
        $1\rightarrow2$ w/o ancilla              & 0.714 (0.039) & 0.723 (0.040) & 0.762 (0.019) & 0.808 (0.024) & & 0.824 (0.024)      \\
        \phantom{$1\rightarrow2$} w/ ancilla   & 0.654 (0.020) & 0.661 (0.019) & 0.690 (0.029) & 0.731 (0.016) & & 0.826 (0.019)     \\
        \midrule
        $1\rightarrow3$ w/o ancilla              & 0.629 (0.024) & 0.635 (0.024) & 0.650 (0.027) & 0.681 (0.028) & & 0.754 (0.023)      \\
        \phantom{$1\rightarrow3$} w/ ancilla   & 0.633 (0.030) & 0.643 (0.029) & 0.613 (0.027) & 0.629 (0.030) & & 0.765 (0.022)      \\
        \bottomrule
    \end{tabular}
    \caption{Mean fidelity and standard deviation for $1\rightarrow2$ and $1\rightarrow3$ telecloning protocols with and without ancilla, in three teleportation implementations:
    (IBMQ) quantum-controlled operations and deferred measurement, 
    (IBMQ) fourfold increase in circuits with different uncontrolled operations and post selection,
    (Quantinuum) directly with mid-circuit measurement and classical feed-forward control.
    On IBMQ, we also performed measurement error mitigation.
    }
    \label{table:summary_statistics}
\end{figure*}

\paragraph*{Contributions}
Our contributions are threefold. First, we present a novel $1 \rightarrow 3$ quantum telecloning circuit that does not use ancilla qubits. Second, and more importantly, we \emph{unify} this novel telecloning circuit along with all previous $1 \rightarrow 3$ and $1 \rightarrow 2$ quantum telecloning circuits, with and without ancilla, into an algorithm for building explicit gate model circuit descriptions centered on Dicke state preparation \cite{baertschi2019deterministic,9774323,baertschi2022shortdepth,PhysRev.93.99,aulbach2010maximally,G_hne_2008}. Additionally, this algorithm \emph{extends} the telecloning circuits with ancilla to $1 \rightarrow M$ clones. This type of circuit model description has not been presented before, instead previous work described $1 \rightarrow M$ telecloning with ancilla only in mathematical terms.

Finally, we provide experimental fidelity measurements from executing these telecloning circuits on cloud accessible NISQ \cite{Preskill2018quantumcomputingin} computers using parallel single qubit tomography. Notably, in these experimental implementations we utilize three different methods for implementing the teleportation component of the telecloning protocol on the devices; \textbf{i)} mid circuit measurement with real time conditional operations on the Quantinuum H1-2 device (this allows for the telecloning protocol to be directly executed in the circuit), \textbf{ii)} classical post selection on the measurements and \textbf{iii)} deferred measurement, where the conditional operations are carried out using CNOT gates in the circuits. Options \textbf{ii)} and \textbf{iii)} are applied when \textbf{i)} is not available, which is the case on the IBMQ devices at the time of these experiments. 

Quantum cloning, and by extension variants including quantum telecloning, are of interest for improvement of some quantum computations \cite{galvao2000cloning}, and to distribute quantum information \cite{ricci2005separating, yang2021experimental}. Within a quantum network, quantum telecloning specifically could allow for transmission of approximate quantum information using classical communication. Beyond the potential use of our implementation and analysis on NISQ devices, our practical implementations include relevant benchmarks in terms of the measured clone fidelity. Our work presents a natural use case for mid circuit measurement with real time classical condition control on NISQ devices. 

All data, code, and extra figures are available on a public Github repository\footnote{\url{https://github.com/lanl/Quantum-Telecloning}}.
An earlier pre-feasibility study of IBMQ experiments is available as a master's thesis~\cite{garcia2022quantum}. Figures in this article were generated using Qiskit~\cite{Qiskit} and Matplotlib~\cite{Hunter:2007, thomas_a_caswell_2021_5194481}. 
Our experimental findings are summarized in the table in Figure~\ref{table:summary_statistics}, with details given in Section~\ref{sec:results}.

%

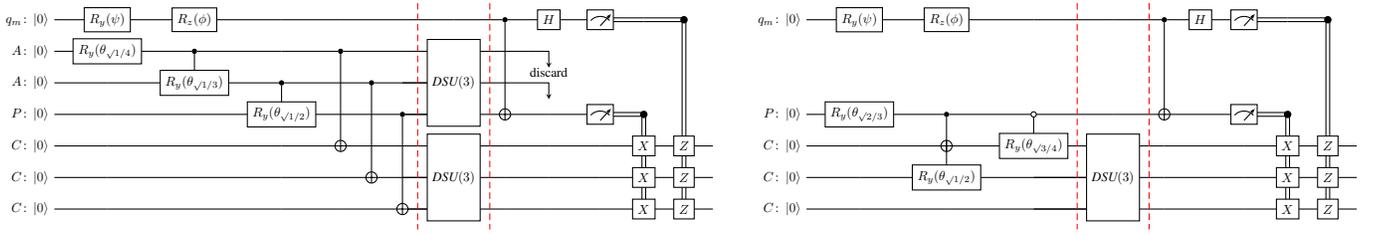
\begin{figure*}[t!]
	\centering
	\begin{adjustbox}{width=\linewidth}
		\begin{quantikz}[row sep={24pt,between origins},execute at end picture={}]
			\lstick{$q_m\colon\ket{0}$}		& \gate{R_y(\psi)}	& \gate{R_z(\phi)}	& \qw		& \qw		& \qw		& \qw\slice{}	& \qw\slice{}		& \ctrl{3}	& \gate{H}		& \meter{}		& \cw		& \cwbend{6}	&	&	&	&	&	& \lstick{$q_m\colon\ket{0}$}		& \gate{R_y(\psi)}	& \gate{R_z(\phi)}	& \qw\slice{}		& \qw\slice{}		& \ctrl{3}	& \gate{H}	& \meter{}	& \cw		& \cwbend{6}	& 	\\
			\lstick{$A\colon \ket{0}$}	& \rygate{1}{4}		& \ctrl{1}		& \qw		& \ctrl{3}	& \qw		& \qw		& \gate[3]{\dsu{3}}	& \qw		& \trash{\text{discard}}& 			&		&		&	&	&	& 	& 	&				&			& 			& 			& 			& 		& 		& 		& 		& 		& 	\\ 
			\lstick{$A\colon \ket{0}$}	& \qw			& \rygate{1}{3}		& \ctrl{1}	& \qw		& \ctrl{3}	& \qw		& \qw			& \qw		& \trash{}		& 			&		&		&	&	&	& 	& 	&				&			& 			& 			& 			& 		& 		& 		& 		& 		& 	\\
			\lstick{$P\colon \ket{0}$}	& \qw			& \qw			& \rygate{1}{2}	& \qw		& \qw		& \ctrl{3}	& \qw			& \targ{}	& \qw			& \meter{}		& \cwbend{3}	&		&	&	&	& 	& 	& \lstick{$P\colon\ket{0}$}	& \rygate{2}{3}		& \ctrl{2}		& \octrl{1}		& \qw			& \targ{}	& \qw		& \meter{}	& \cwbend{3}	& 		& 	\\
			\lstick{$C\colon \ket{0}$}	& \qw			& \qw			& \qw		& \targ{}	& \qw		& \qw		& \gate[3]{\dsu{3}}	& \qw		& \qw			& \qw			& \gate{X}	& \gate{Z}	& \qw	&	&	& 	& 	& \lstick{$C\colon\ket{0}$}	& \qw			& \targ{}		& \rygate{3}{4}		& \gate[3]{\dsu{3}}	& \qw		& \qw		& \qw		& \gate{X}	& \gate{Z}	& \qw	\\
			\lstick{$C\colon \ket{0}$}	& \qw			& \qw			& \qw		& \qw		& \targ{}	& \qw		& \qw			& \qw		& \qw			& \qw			& \gate{X}	& \gate{Z}	& \qw	&	&	& 	& 	& \lstick{$C\colon\ket{0}$}	& \qw			& \rygate{1}{2}		& \qw			& \qw 			& \qw		& \qw		& \qw		& \gate{X}	& \gate{Z}	& \qw	\\
			\lstick{$C\colon \ket{0}$}	& \qw			& \qw			& \qw		& \qw		& \qw		& \targ{}	& \qw			& \qw		& \qw			& \qw			& \gate{X}	& \gate{Z}	& \qw	&	&	& 	& 	& \lstick{$C\colon\ket{0}$}	& \qw			& \qw			& \qw			& \qw			& \qw		& \qw		& \qw		& \gate{X}	& \gate{Z}	& \qw			
		\end{quantikz}
	\end{adjustbox}

	\caption{Two quantum telecloning circuits for three clones, with and without the use of ancilla for the telecloning states. We use little-endian ket notation (top-to-bottom wires correspond to left-to-right bitstrings) and shorthand angle notation $\ryangle{x}{y} = 2\cos^{-1}(\sqrt{x/y})$ such that $R_y(\ryangle{x}{y})\ket{0} = \surd\tfrac{x}{y}\ket{0} + \surd\tfrac{y-x}{y}\ket{1}$:\newline
	\emph{(left)} Using an AAPCCC telecloning state (2 ancilla qubits $A$, 1 port qubit $P$, $M=3$ clone qubits $C$): First prepare the message qubit on the top wire, and the state $\tfrac{1}{\smash{\surd{M}}}\sum_{i=0}^M \ket{1^i 0^{M-i}}\ket{1^i 0^{M-i}}$. Symmetrize $AAP$ and $CCC$ using two Dicke state unitaries $\dsu{M}$ to get $\tfrac{1}{\smash{\surd{M}}} \sum_{i=0}^M \dicke{M}{i}\dicke{M}{i}$. Discard the ancilla qubits, perform a Bell measurement on the message and port qubits, and classically communicate the results to the holders of the clone qubits for them to control Pauli-X and Pauli-Z adjustments, similar to quantum teleportation.\newline
	\emph{(right)} Using a PCCC telecloning state instead (1 port $P$, $M=3$ clones $C$): Prepare 
	$\smash{\surd{\tfrac{2}{3}}} \ket{0} \left( \smash{\surd{\tfrac{3}{4}}} \ket{000} + \smash{\surd{\tfrac{1}{4}}} \ket{100} \right) + \smash{\surd{\tfrac{1}{3}}} \ket{1} \left( \smash{\surd{\tfrac{1}{2}}} \ket{100} + \smash{\surd{\tfrac{1}{2}}} \ket{110} \right)$.
	Symmetrize the clone qubits $CCC$ using a single Dicke state unitary $\dsu{3}$ to get the state $\smash{\surd{\tfrac{1}{2}}}\ket{0}\dicke{3}{0} + \smash{\surd{\tfrac{1}{6}}}\ket{0}\dicke{3}{1} + \smash{\surd{\tfrac{1}{6}}}\ket{1}\dicke{3}{1} + \smash{\surd{\tfrac{1}{6}}}\ket{1}\dicke{3}{2}$. Perform a Bell measurement on the message and port qubits, followed by classical communication to and local Pauli operations on the clone qubits. 
	}
	\label{fig:main-circuit}
\end{figure*}

\section{Methods}
\label{sec:methods}
In this section we first detail the algorithmic circuit construction for the quantum telecloning circuits. This includes the circuit construction using different hardware connectivities. In Section \ref{sec:methods_circuit_and_telecloning_protocol} we detail the three different implementations of quantum telecloning; deferred measurement, post selection, and mid-circuit measurement with real time conditional operations. Lastly, in Section \ref{sec:methods_workflow} we describe the state tomography estimation method used and the hardware dependent circuit execution parameters used.

\subsection{Quantum Telecloning Protocol and Circuits}
\label{sec:methods_circuit_and_telecloning_protocol}
We begin by noting some key features of the quantum telecloning protocol. The teleportation component of telecloning enables the transmission of quantum information through a Bell state \cite{Gisin_1998} measurement. The four Bell states $\Phi^+, \Phi^-, \Psi^+, \Psi^-$ are 2-qubit states that form a maximally entangled basis, here defined as $\ket{\text{port,message}}$:
\begin{align*}
    \Phi^{\pm}\colon \frac{1}{\sqrt{2}}(\ket{00} \pm \ket{11}), 
    &\qquad \Psi^{\pm}\colon \frac{1}{\sqrt{2}}(\ket{01} \pm \ket{10}) 
\end{align*}

\begin{algorithm}[b!]
\caption{Quantum telecloning protocol for cloning of $N=1$ quantum state to $M > N$ clones}
\begin{algorithmic}[1]
\Statex \hspace*{-2.5ex}\textbf{State Preparation:}
\State A message qubit $q_m$ is prepared by a sender
\State A quantum telecloning state $TC$ is constructed with
\Statex -~$1$ Port qubit, and potentially $M-1$ ancilla qubits.
\Statex -~$M$ clone qubits (sent to the receivers) plus
\Statex $TC$ is symmetric in the clones, and symmetric between the port and ancilla qubits (i.e. they are interchangeable).
\Statex \hspace*{-2.5ex}\textbf{Teleportation:}
\State A bell measurement is made between $q_m$ and the Port qubit of $TC$, and the results are communicated classically to the clone holders.
\State The clone holders use the result of the bell measurement to decide whether to apply $X$- and/or $Z$-gates to the clone qubits in order to construct the approximate clones:
\Statex - $\Phi^+$: apply nothing
\Statex - $\Phi^-$: apply $Z$-gate
\Statex - $\Psi^+$: apply $X$-gate
\Statex - $\Psi^-$: apply $X$- then $Z$-gate
\Statex \hspace*{-2.5ex}\textbf{Result:}
\State $M$ approximate clones of $q_m$ have been generated with theoretical maximal fidelity described by Equation \ref{eq:theoretical-fidelity}.
\end{algorithmic}
\label{algorithm:telecloning}
\end{algorithm}

\noindent Another important yet subtle feature of the quantum telecloning protocol (see Algorithm \ref{algorithm:telecloning}) is that the message qubit, the telecloning state, and the resulting clones do not need to be spatially connected at all times. In particular, the relevant operation is the bell measurement on the message qubit by the telecloning state which subsequently generates the clones, meaning that the message qubit does not need to be initialized within the same proximity as the telecloning state. Another principal feature is the structure of the telecloning state, which plays two key roles, it produces optimal clones, and acts as a quantum channel to distribute the quantum information. We wish for this distribution of information to be symmetric between the clones, therefore, a fitting choice for this telecloning state are Dicke states, which contain the necessary entanglement and are symmetric under permutation of qubits (e.g., port and ancilla qubits) \cite{Burchardt_2021}. We construct telecloning states (illustrated in a quantikz~\cite{quantikz} circuit in Figure \ref{fig:main-circuit} for three clones with and without ancilla) using Dicke state unitaries $\dsu{M}$ \cite{baertschi2019deterministic, 9774323}, defined as unitaries mapping any input Hamming weight $i$ in unary encoding to the Dicke state \(\dicke{M}{i}\):
\[ \forall 0\leq i \leq M\colon \dsu{M}\colon \ket{1^i 0^{M-i}} \mapsto \dicke{M}{i}, \]
where the Dicke states \(\dicke{M}{i}\) are a superposition of bitstrings $x$ of length $M$ of Hamming weight (number of ones) $hw(x) = i$:
\[ \dicke{M}{i} = \tbinom{M}{i}^{-1/2} \sum\nolimits_{x\in\{0,1\}^M,\ hw(x)=i} \ket{x} \]
For $1 \rightarrow M$ telecloning with ancilla, the telecloning state is composed of $M-1$ ancilla qubits, $1$ port qubit, and $M$ clone qubits, given by:
\[  \sqrt{\tfrac{1}{M}} \sum\nolimits_{i=0}^M \dicke{M}{i} \dicke{M}{i} \]
For $1 \rightarrow 2$ telecloning circuits without ancilla, the state is composed of 1 port qubit and 2 clone qubits:
\[ \surd{\tfrac{2}{3}} \ket{0}\dicke{2}{0} + \surd{\tfrac{1}{3}} \ket{1}\dicke{2}{1} \]
Lastly, for $1 \rightarrow 3$ telecloning circuits without ancilla, the state consists of 1 port qubit and 3 clone qubits, novel for symmetric universal telecloning:
\[ \surd{\tfrac{1}{2}}\ket{0}\dicke{3}{0} + \surd{\tfrac{1}{6}}\ket{0}\dicke{3}{1} + \surd{\tfrac{1}{6}}\ket{1}\dicke{3}{1} + \surd{\tfrac{1}{6}}\ket{1}\dicke{3}{2} \]

\begin{figure*}[t!]
    \centering
    \includegraphics[width=1\textwidth]{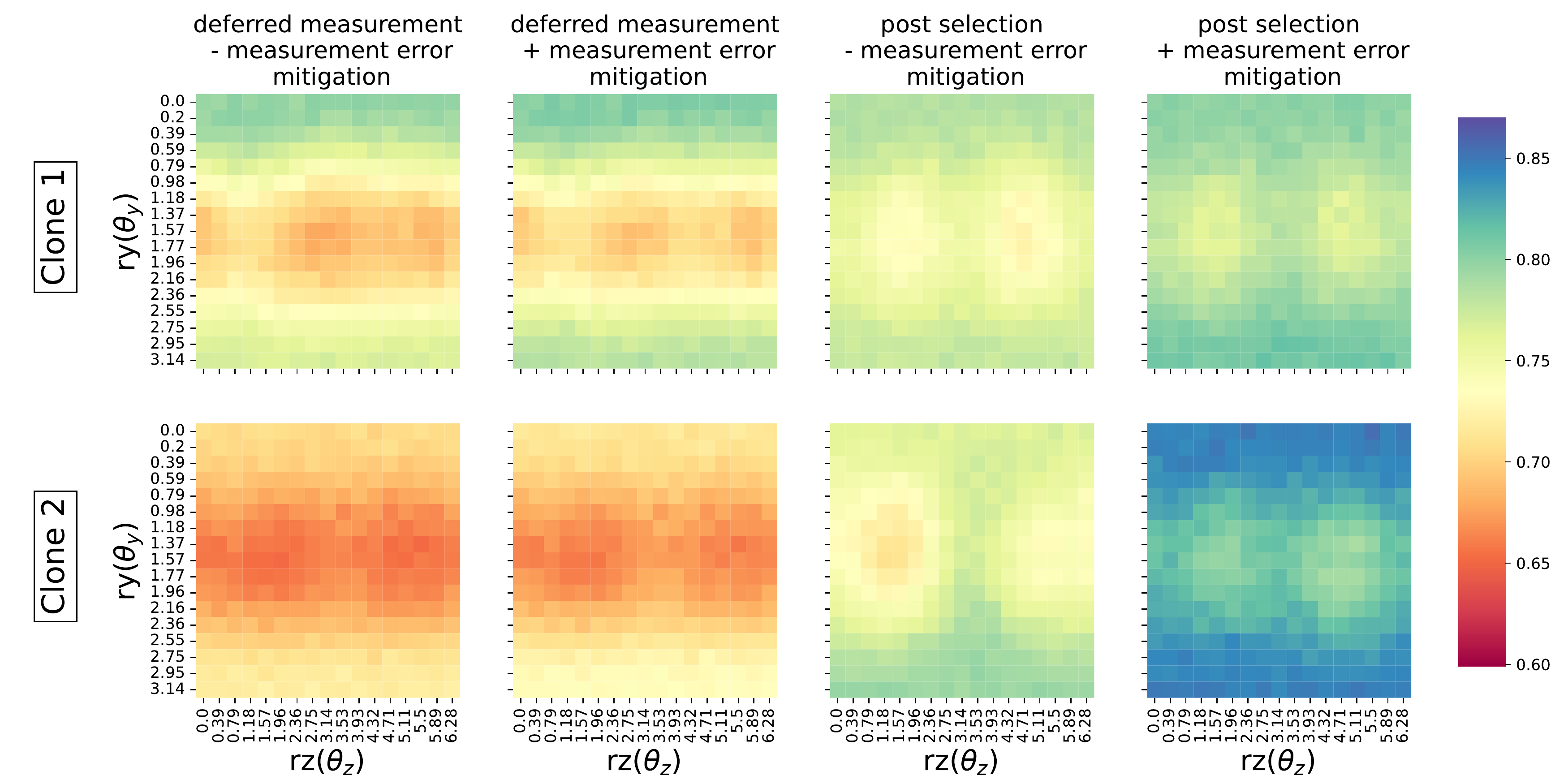}
    \caption{PCC circuit implemented in the form of deferred measurement and post selection of ibmq\_montreal. The theoretical maximum fidelity for two clones is $0.8\overline{3}$. First and third column show the raw fidelity results without measurement error mitigation; the second and fourth columns show the same results with measurement error mitigation applied. The mean clone fidelity for deferred measurement without measurement error mitigation was $0.714$ with a standard deviation (SD) of $0.039$. The mean clone fidelity for deferred measurement with measurement error mitigation was $0.723$ with SD $0.040$. The mean clone fidelity for post selection without measurement error mitigation was $0.762$ with SD $0.019$. The mean clone fidelity for post selection with measurement error mitigation was $0.808$ with SD $0.024$. Executed on qubits \textbf{0, 1, 4, 7} on ibmq\_montreal. The qubit clone indices, for the logical circuit, are $0$, $2$ for PCC deferred measurement and $2$, $3$ for PCC using post selection. The numbering on the rows corresponds to this order (i.e., the clone number is index based).}
    \label{fig:PCC}
\end{figure*}

In order to emulate the quantum telecloning protocol on a NISQ computer, the important mechanism is how to use the bell measurement (classical) information to generate the clones. In order to do this the NISQ device would need to execute \emph{mid-circuit measurement with subsequent classical control features} (i.e. if statements) to execute X/Z Pauli gates which then create approximate clones of the message qubit. This feature is available on the Quantinuum H1-2 device \cite{pino2021demonstration}, which allows direct emulation of the quantum telecloning protocol. On the IBMQ hardware this is not currently supported. Therefore we present two alternative methods: \emph{deferred measurement} and \emph{post selection}. 

The first alternative method, \emph{deferred measurement}, is where instead of using the classical control to make decisions on what quantum operations to then apply, all of the control is done using quantum gates \cite{nielsen2000quantum}. This has the advantage of removing the requirement of using classical feed-forward control operations, which are not yet available on many NISQ devices. The primary disadvantage of this mechanism is that it costs more CNOT's to implement.

The second alternative method, \emph{post selection}, is where the Pauli X and/or Z gates are applied a priori (i.e. to all circuits) during the circuit execution; and then subsequently all measured states that do not correspond to that quantum control logic are removed during post processing. This is then repeated for all of the possible measured classical states of the port and message qubits (i.e. $11$, $10$, $01$, $00$). Therefore, $4$ separate quantum circuits are executed (for a given message qubit) in order to implement post selection, and each of those runs will discard many of the measured results. The primary downside with this method is that it requires the execution of $4$ separate circuits for each telecloning circuit.

We define the telecloning circuit names using the following convention: \textbf{P} indicates the port qubit, \textbf{A} indicates an ancilla qubit, and \textbf{C} indicates a clone qubit. Thus, a combination of these letters encodes the circuit construction; for example AAPCCC indicates a telecloning circuit which uses two ancilla, one port, and creates three clone qubits. Note that in this naming convention, the \emph{message} qubit is implicitly included because it is the state that is cloned. 

We give circuit constructions for two underlying NISQ computer architectures; full all-to-all connectivity (Quantinuum compatible) and LNN connectivity between qubits of the Telecloning state + edge between Message \& Port qubits (IBMQ compatible). Quantinuum H1-2 has 12 qubits (at the time of the experiments), and the ibmq\_montreal has 27 qubits; thus both devices fit the telecloning circuit requirements. 

Among the available quantum operations on NISQ devices, two-qubit gates and measurements have the highest error rates. Therefore, we now summarize CNOT costs of our circuits in terms of the number of clones $M$, for telecloning states with and without ancilla, see Figure~\ref{fig:main-circuit} (left) and (right), respectively:

\emph{State Preparation Cost}:
In order to prepare a telecloning state with ancilla, the cost \emph{before} symmetrization with Dicke state unitaries $\dsu{M}$ is $2M^2-M$ CNOTs on LNN connectivity and $2M -1$ CNOTs on all-to-all connectivity. 
For the telecloning state without ancilla this cost is 1 CNOT for two clones (regardless of connectivity), $2$ CNOTs for three clones on full connectivity, and $3$ CNOTs for three clones on LNN connectivity.
Next, the Dicke state unitary preparation costs $2.5 M^2 - 5.5M + 3$ per unitary \cite{9774323}. With ancilla, the preparation requires two $\dsu{M}$ unitaries whereas without ancilla it requires only a single Dicke state unitary. 

\emph{Teleportation Cost}:
The bell measurement always uses exactly one CNOT gate across all circuits, 
if implemented with mid-circuit measurement with real time classical feed-forward conditionals. The same holds for our post-selection circuits (albeit at the overhead of running four different circuits).
However,
\emph{Deferred measurement} requires additional CNOT gates, adding $4M-1$ CNOTs on LNN connectivity, and $2M$ CNOTs on full connectivity (a setting we do not use). 

Thus using the template in Figure \ref{fig:main-circuit} the CNOT cost of any of our implementations of quantum telecloning circuits can be computed in a straightforward fashion: As an example, the CNOT cost of AAPCCC ($M=3$ clones with ancilla) when implemented with deferred measurement on LNN connectivity is: $2M^2 - M + 2\cdot (2.5 M^2 - 5.5 M + 3) + 1 + 4M - 1 = 45$. 

\begin{figure*}[t!]
    \centering
    \includegraphics[width=1\textwidth]{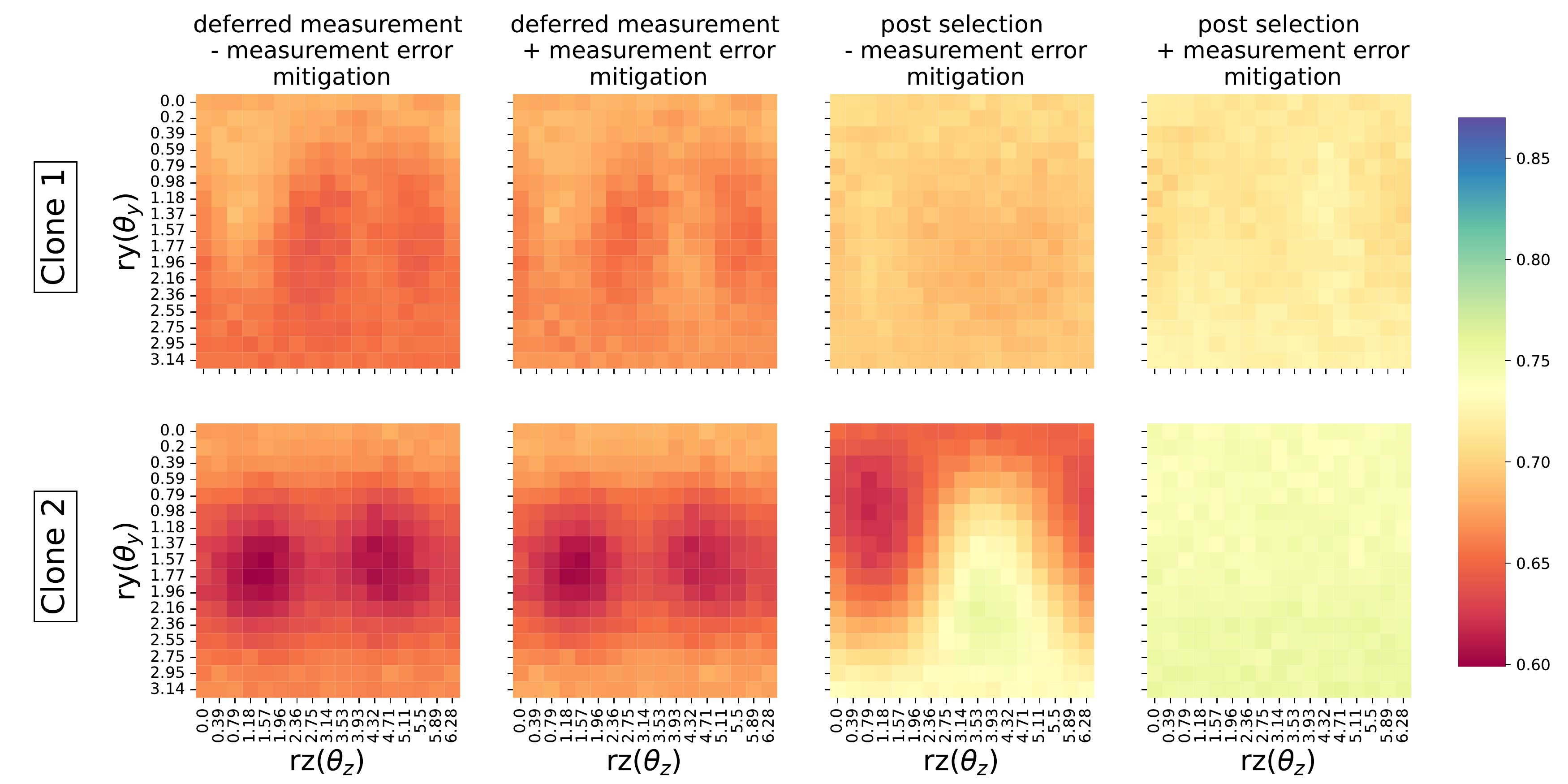}
    \caption{APCC circuit implemented in the form of deferred measurement and post selection of ibmq\_montreal. The theoretical maximum fidelity for two clones is $0.8\overline{3}$. First and third column show the raw fidelity results without measurement error mitigation, and the second and fourth columns show the same results with measurement error mitigation applied. The mean clone fidelity for deferred measurement without measurement error mitigation was $0.654$ with SD $0.020$. The mean clone fidelity for deferred measurement with measurement error mitigation was $0.661$ with SD $0.019$. The mean clone fidelity for post selection without measurement error mitigation was $0.690$ with SD $0.029$. The mean clone fidelity for post selection with measurement error mitigation was $0.731$ with SD $0.016$. Executed on qubits \textbf{0, 2, 1, 4, 7} on ibmq\_montreal. The qubit clone indices, for the logical circuit, are $0$, $3$ for APCC deferred measurement and $3$, $4$ for APCC post selection. The clone numbering on the rows corresponds to this numbering order (in other words the clone number is index based). }
    \label{fig:APCC}
\end{figure*}

\begin{figure*}[t!]
    \centering
    \includegraphics[width=1\textwidth]{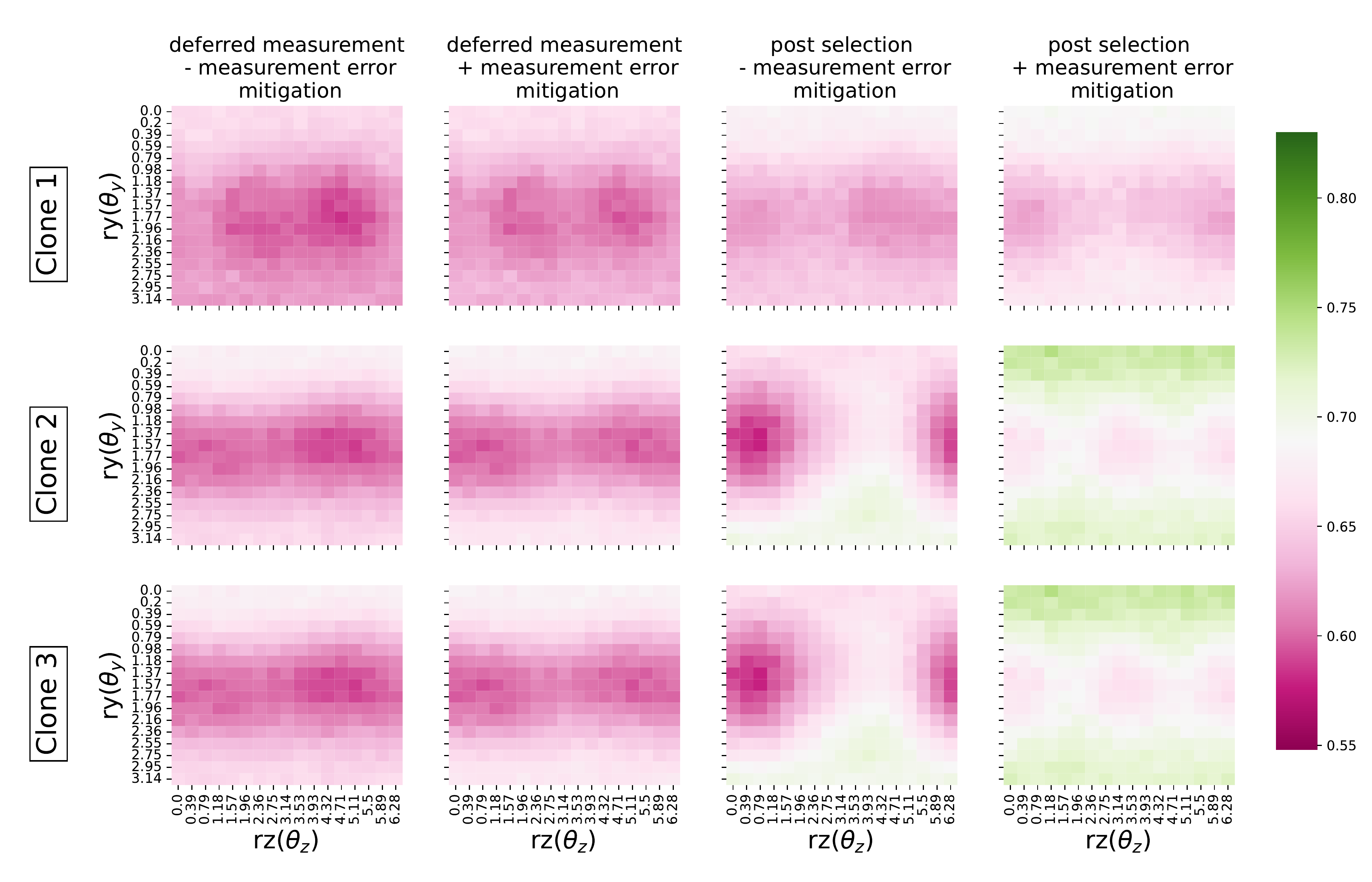}
    \caption{PCCC circuit implemented in the form of deferred measurement and post selection of ibmq\_montreal. The theoretical maximum fidelity for three clones is $0.\overline{7}$. First and third column show the raw fidelity results without measurement error mitigation, and the second and fourth columns show the same results with measurement error mitigation applied. The mean clone fidelity for deferred measurement without measurement error mitigation was $0.629$ with SD $0.024$. The mean clone fidelity for deferred measurement with measurement error mitigation was $0.635$ with SD $0.024$. The mean clone fidelity for post selection without measurement error mitigation was $0.650$ with SD $0.027$. The mean clone fidelity for post selection with measurement error mitigation was $0.681$ with SD $0.028$. Executed on qubits \textbf{0, 1, 4, 7, 10} on ibmq\_montreal. The qubit clone indices, for the logical circuit, are $0$, $1$, $3$ for PCCC deferred measurement and $2$, $3$, $4$ for PCCC post selection. The numbering on the rows corresponds to this numbering order (i.e., the clone number is index based). }
    \label{fig:PCCC}
\end{figure*}

Lastly, we give interactive Quirk circuits 
\cite{quirk} for our $1\rightarrow 2$ Telecloning and $1\rightarrow 3$ Telecloning circuits (with and without ancilla). In these circuits the state of the message qubit is being varied over time. Note that with ancilla, the clones' mixed state \& fidelity is independent from the outcome of the Bell measurement of the Message \& Port qubits, and without ancilla, the clones mixed state \& fidelity is averaged over different mixed states \& fidelities depending on the outcome from the Bell measurements. The outcome of the SWAP test relates to the fidelity measure as $(\text{fidelity})=1-2(\text{ON percentage})$:
\href{https://algassert.com/quirk#circuit=
}{$1\rightarrow 2$ Telecloning Variants},
\href{https://algassert.com/quirk#circuit=
}{$1\rightarrow 3$ Telecloning Variants}.

\subsection{Hardware implementation details}
\label{sec:methods_workflow}
In this section the workflow of analyzing and executing the quantum telecloning implementations is described. 
Quantum state tomography in general can be computationally intensive, however in this case we are only performing state tomography on single qubits. In order to reduce Quantum Processing Unit (QPU) time usage, we implement parallel single qubit state tomography \cite{PhysRevLett.124.100401} to measure the fidelity of all of the clones. Thus the required number of circuits to compute the state tomography of $M$ clone qubits is always $3$ (one for each of the three Pauli basis states X/Y/Z). In order to compute the density matrices of the clones, we use the maximum likelihood estimation quantum tomography fitter that is implemented, with slight modifications, in Qiskit Ignis \cite{Qiskit, PhysRevLett.108.070502}. Once the density matrices of the clones have been constructed, we compute exactly the original density matrix of the message qubit given the known single qubit gates we have applied to initialize the message qubit in the experiments. Then we use Equation \ref{eq:fidelity-general} in order to compute the \emph{fidelity}~\cite{jozsa1994fidelity} of the clone(s) in comparison to the pure quantum state of the message qubit. 
\begin{equation}
    F(\rho_1, \rho_2) = Tr[ \sqrt{ \sqrt{\rho_1}\rho_2 \sqrt{\rho_1} } ]^2
    \label{eq:fidelity-general}
\end{equation}
Note that since the message qubit is a pure state $\rho_1 = \sqrt{\rho_1} = \ket{\psi_1}\bra{\psi_1}$, Equation \ref{eq:fidelity-general} simplifies to Equation \ref{eq:fidelity-single-pure-state}:
\begin{equation}
    F(\rho_1, \rho_2) = \braket{\psi_1 | \rho_2 | \psi_1}
    \label{eq:fidelity-single-pure-state}
\end{equation}

In practice we need to estimate the state of the clones using state tomography, and the combination of the least squares estimation process and statistical noise due to a finite number of samples, can result in fidelities which are slightly larger than the theoretical limit given by Equation \ref{eq:theoretical-fidelity}. 

Importantly, for all three variants of the quantum telecloning circuits, the delineation between \textbf{1)} the telecloning state construction, \textbf{2)} the Bell measurement on the message qubit by the telecloning state, and lastly \textbf{3)} the parallel state tomography is essential. In order to encode this separation between the components of the telecloning circuits, the \textbf{barrier} operation is used. Specifically, two barriers are always present to separate the three telecloning protocol components, although in some cases the measurements are also immediately preceded by a third barrier. 

\begin{figure*}[t!]
    \centering
    \includegraphics[width=1\textwidth]{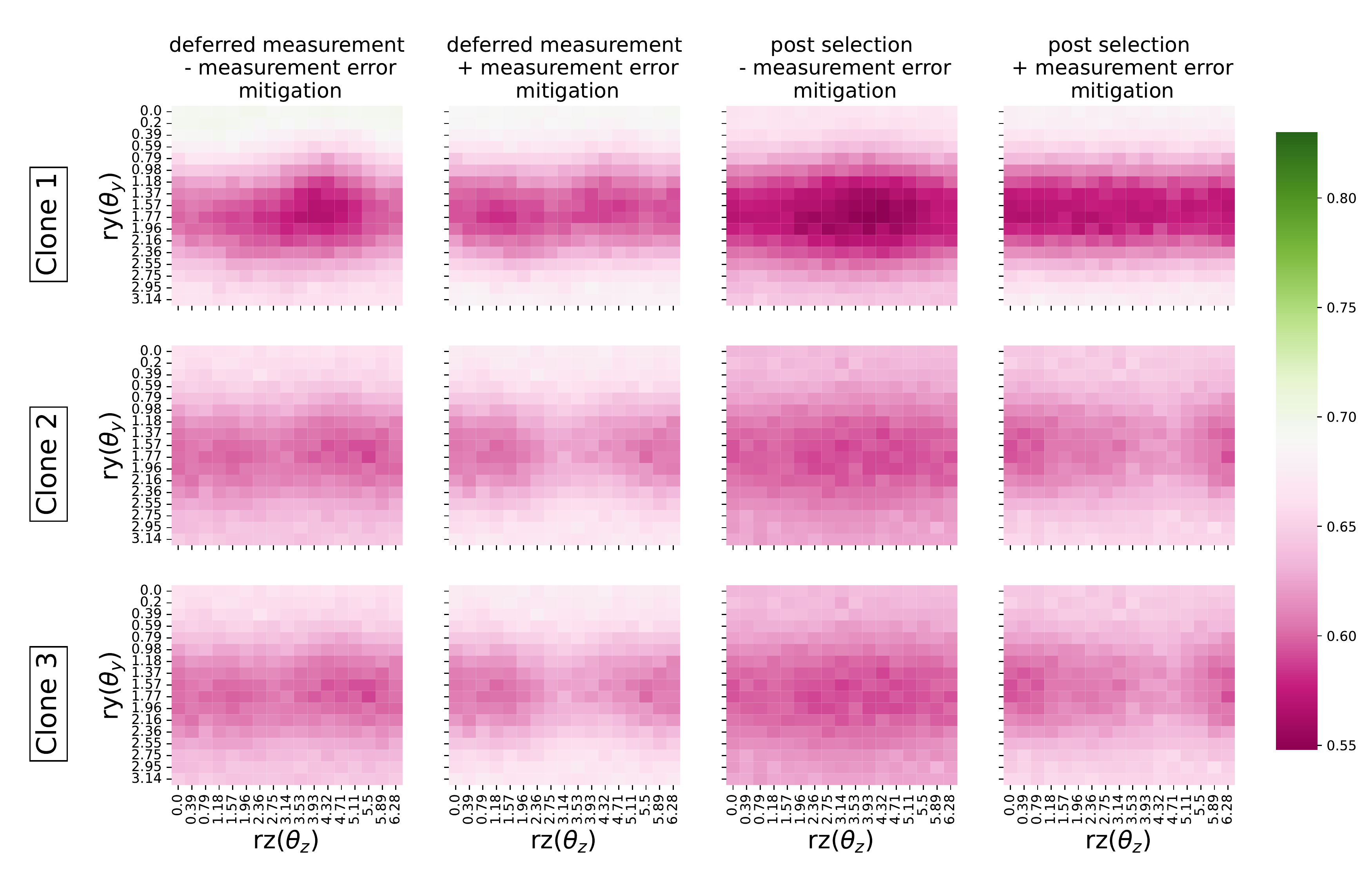}
    \caption{AAPCCC circuit implemented in the form of deferred measurement and post selection of ibmq\_montreal. The theoretical maximum fidelity for three clones is $0.\overline{7}$. First and third column show the raw fidelity results without measurement error mitigation, and the second and fourth columns show the same results with measurement error mitigation applied. The mean clone fidelity for deferred measurement without measurement error mitigation was $0.633$ with SD $0.030$. The mean clone fidelity for deferred measurement with measurement error mitigation was $0.643$ with SD $0.029$. The mean clone fidelity for post selection without measurement error mitigation was $0.613$ with SD $0.027$. The mean clone fidelity for post selection with measurement error mitigation was $0.629$ with SD $0.030$. Executed on qubits \textbf{9, 3, 5, 8, 11, 14, 16} on ibmq\_montreal. The qubit clone indices, for the logical circuit, are $0$, $3$, $5$ for AAPCCC deferred measurement and $4$, $5$, $6$ for AAPCCC post selection. The numbering on the rows corresponds to this order (i.e., the clone number is index based).}
    \label{fig:AAPCCC}
\end{figure*}

In order to improve the clone fidelity, we implement measurement error mitigation on the IBMQ results using Qiskit Ignis 
\cite{Qiskit-Textbook,dewes2012characterization, jattana2020general}. This technique has been utilized in previous studies in order to obtain higher quality results on NISQ devices \cite{alexander2020qiskit, 9774323, wood2020special}. The full measurement error mitigation procedure involves creating and running $2^m$ circuits which are all possible combinations of X gates or Identity gates (i.e. no gates) on each qubit followed by $m$ measurements on the specific qubit subset being addressed. This then gives a full mapping of the intended prepared states to the measured states on the device. It is then possible to use this mapping to remove some of the measurement error from different executed circuits (in our cases quantum telecloning circuits) that have a similar measurement error profile \cite{jattana2020general}. Therefore, ideally this measurement error characterization would be run in sequence with the relevant job that we want to reduce the measurement error on. In our case, we run all varying message qubit state circuits and all measurement error mitigation circuits (for that specific qubit subset) in the same job. The disadvantage of the measurement error mitigation procedure is that it is not scalable to a large number of measured qubits, however for our experiments the largest number of circuit measurements is $5$ (in the cases of post selection for PCCC and AAPCCC) which requires $32$ measurement error mitigation circuits. Measurement error mitigation is not applied to experiments on the Quantinuum H1-2 device because the SPAM error rates on the H1-2 device are much lower than those of the IBMQ devices, and the associated cost with executing full measurement error mitigation on the H1-2 device was not practical (here the primary constraint is the readout time associated with the trapped ion technology compared to the superconducting qubits of IBMQ). 

The Qiskit transpiler \cite{Qiskit} was applied with optimization level $3$ to all of the circuits before they are submitted to the IBMQ backends. Optimization level $3$ is the highest level offered by the transpiler. Besides the circuit optimization in terms of gate depth, the Qiskit transpiler maps the logical circuit qubits onto physical qubits and converts all gates into the IBMQ basis gates \textbf{rz, sx, x, cx}; which prepares the circuits for direct submission to the backend. The use of the Qiskit transpiler did not reduce the CNOT counts in the telecloning circuits (see Section \ref{sec:methods_circuit_and_telecloning_protocol}), although it did reduce the number of single qubit gates after the logical circuits were represented in the IBMQ native gateset. 

\begin{figure*}[t!]
    \centering
    \includegraphics[width=0.49\textwidth]{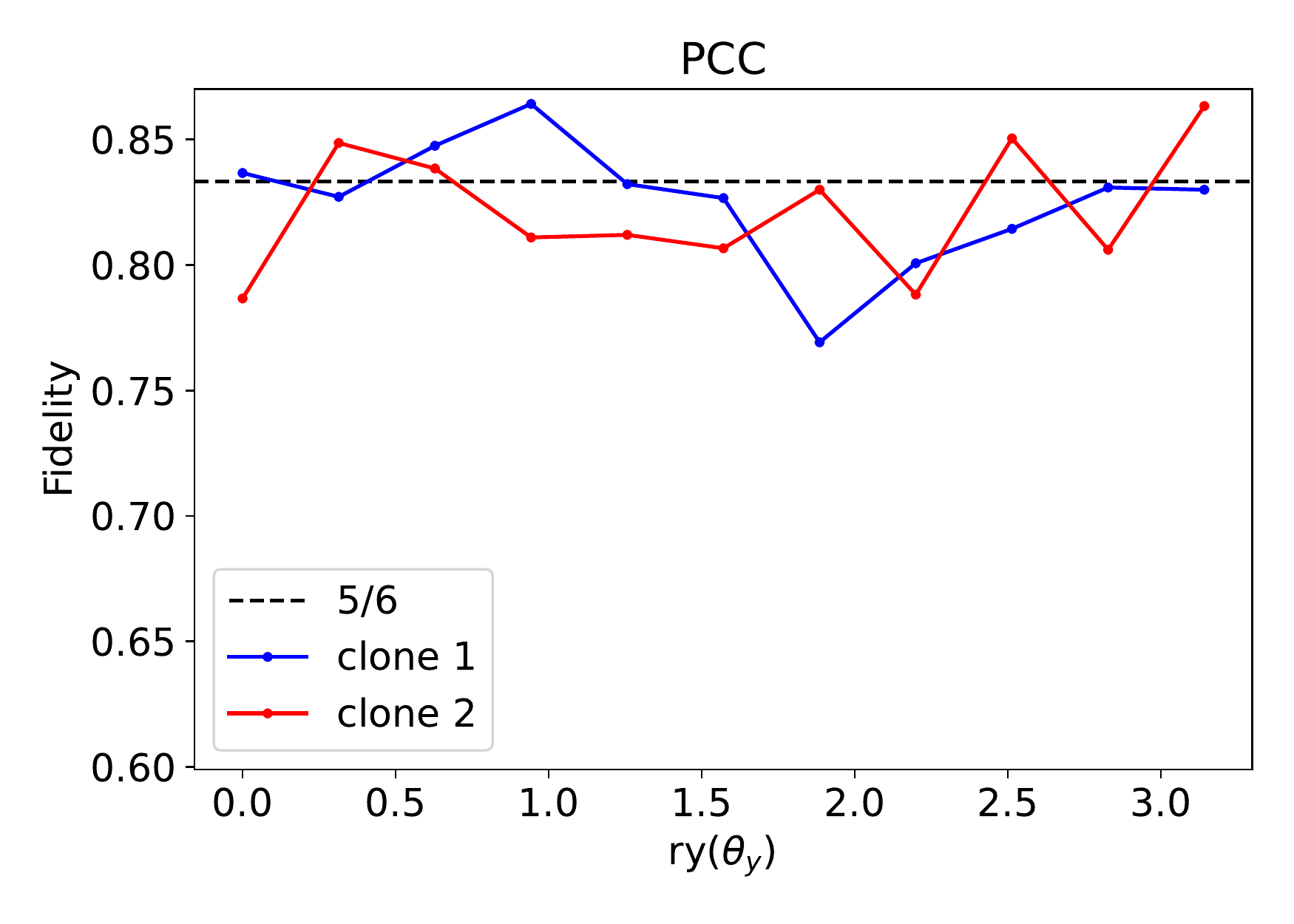}
    \includegraphics[width=0.49\textwidth]{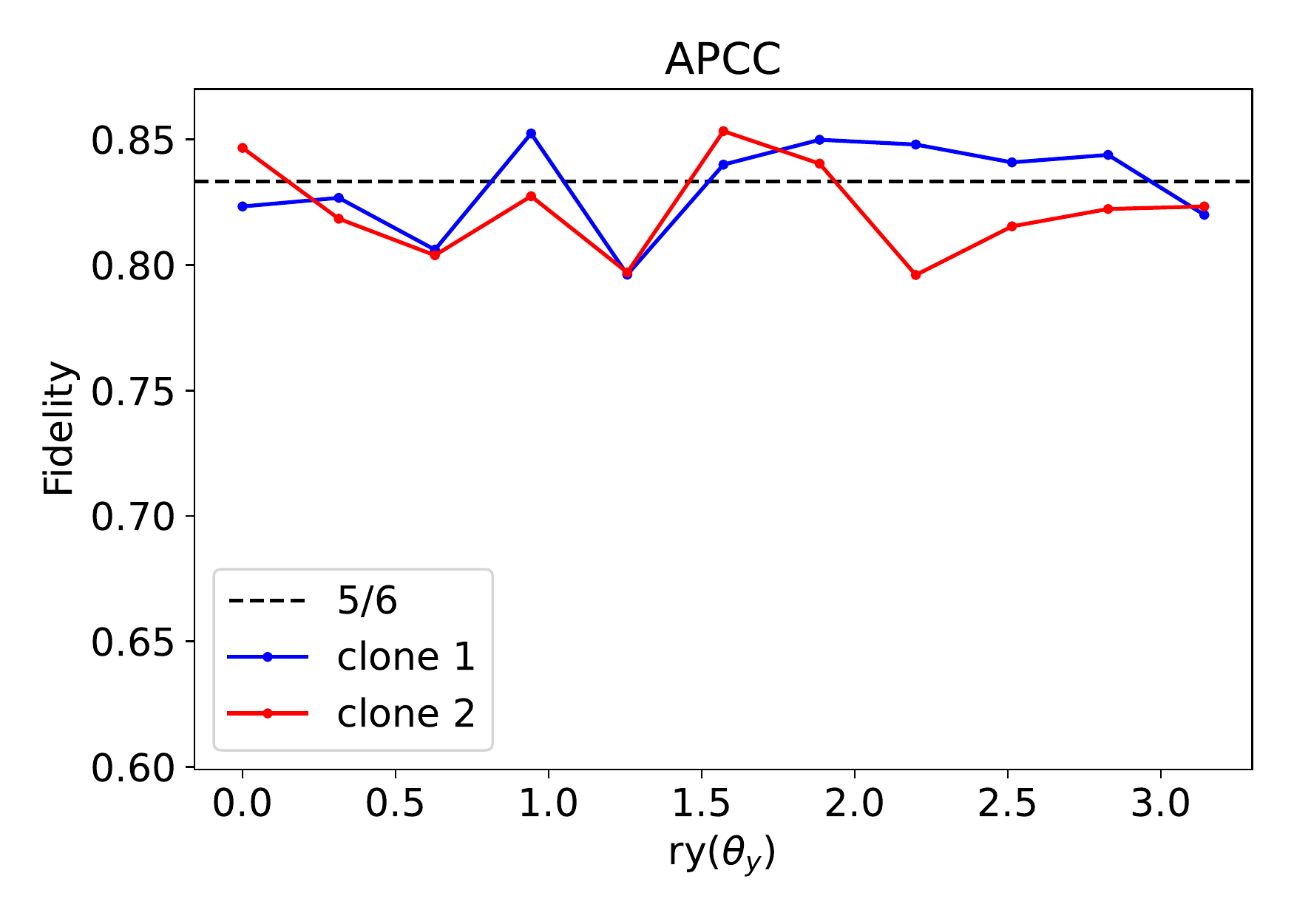}
    \includegraphics[width=0.49\textwidth]{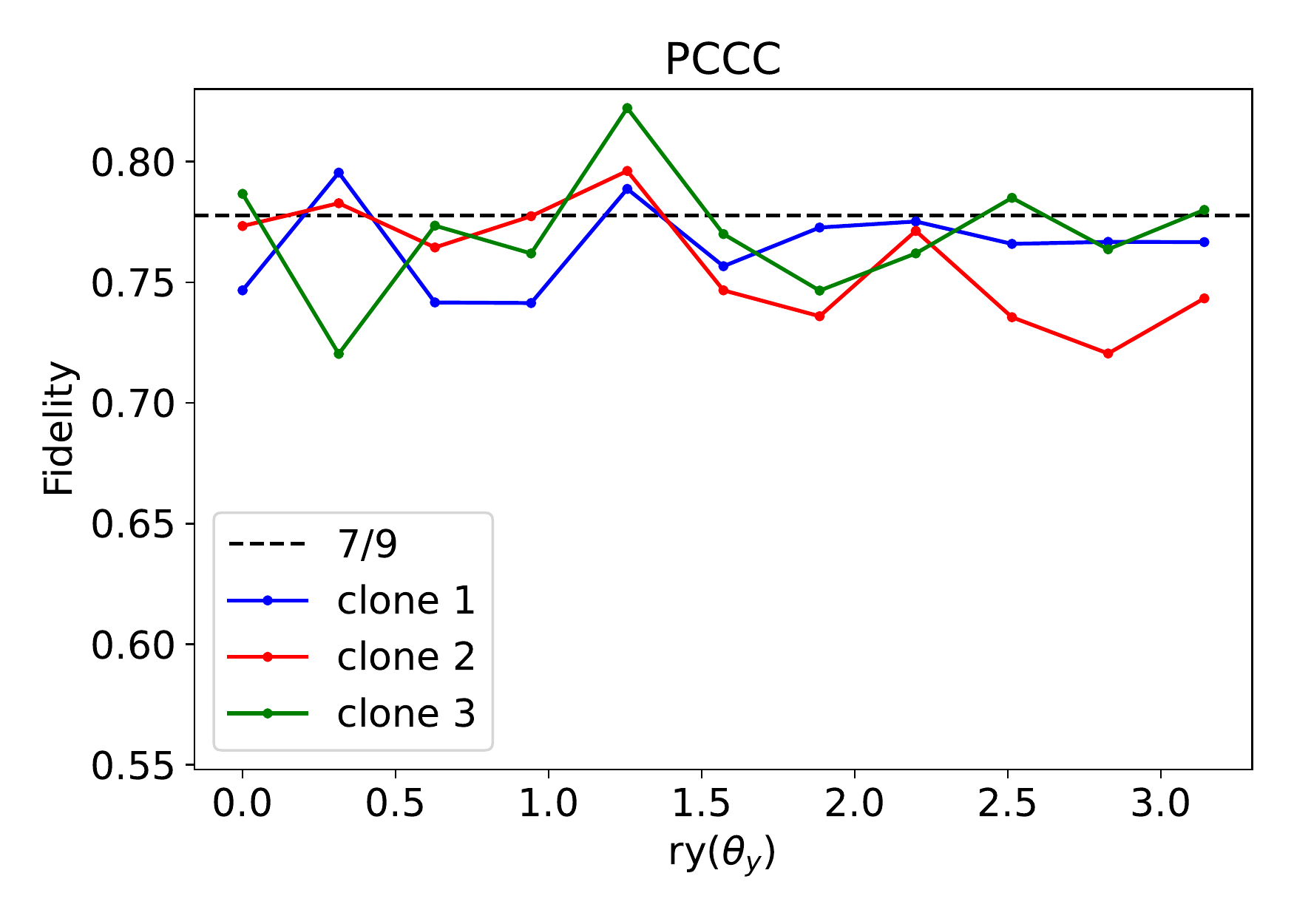}
    \includegraphics[width=0.49\textwidth]{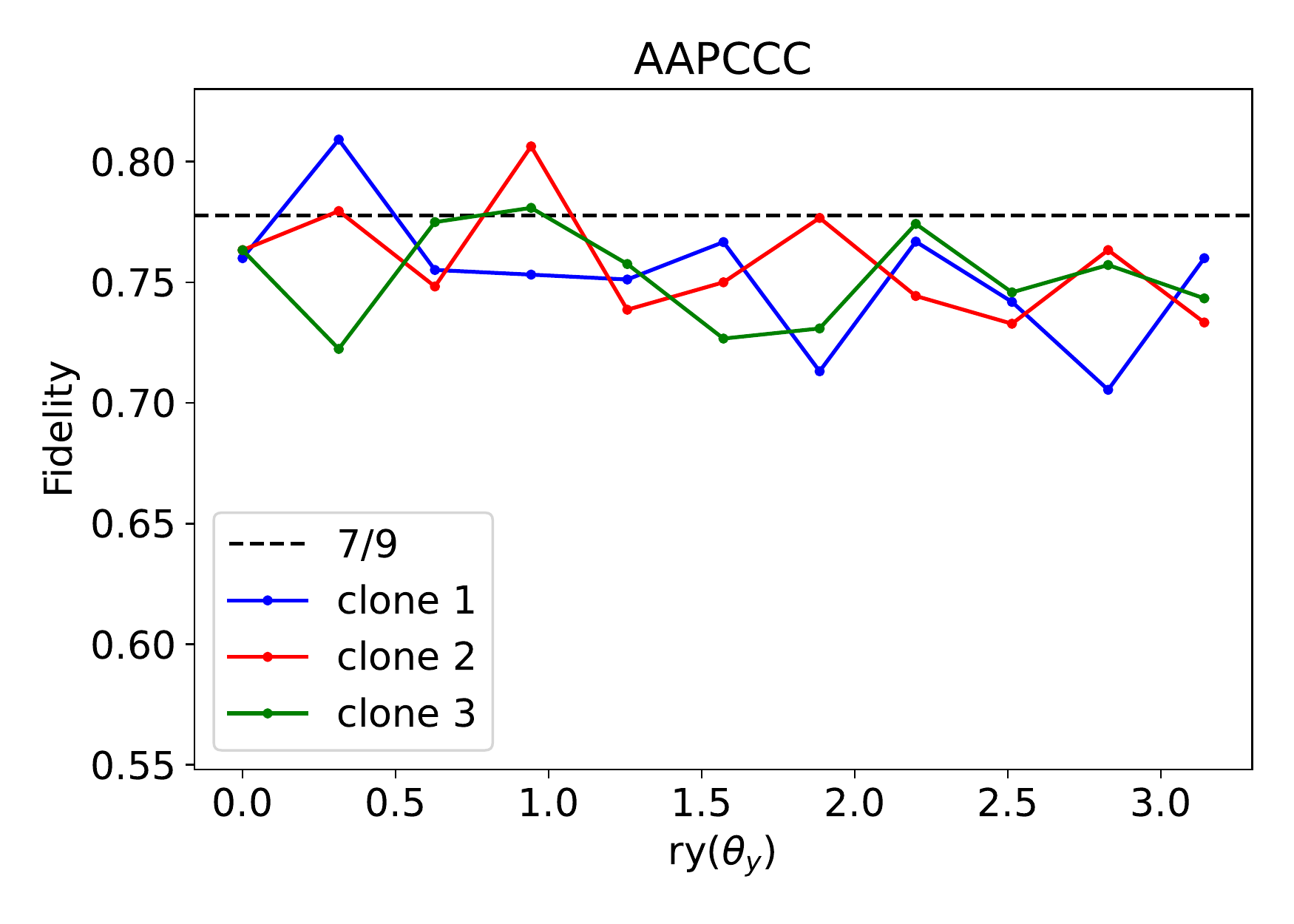}
    \caption{Measured fidelity values as a function of $\theta_y$ ($\theta_z = \frac{\pi}{2}$) across all clones for the PCC circuit (upper left) had a mean of $0.824$ with SD $0.024$,
    for the APCC circuit (upper right) had a mean of $0.826$ with SD $0.019$, 
    for the PCCC circuit (lower left) had a mean of $0.765$ with SD of $0.022$, and
    for the AAPCCC circuit (lower right) had a mean of $0.754$ with a SD of $0.023$.
    These circuits were implemented on the Quantinuum H1-2 with mid-circuit measurement and real time classical if statements. The theoretical maximum fidelity for $2$ and $3$ clones is $0.8\overline{3}$ and $0.\overline{7}$ respectively.}
    \label{fig:Quantinuum_results}
\end{figure*}

The primary characterization we measure is how the fidelity between the pure state of the message qubit and $M$ clones changes as a function of the message qubit state that is being cloned. To this end, the message qubit state is paramaterized by two angles; $\theta_y$ and $\theta_z$, which correspond to the single qubit gates \textbf{ry} and \textbf{rz} applied in sequence (as in Figure \ref{fig:main-circuit}): $ry(\theta_y)$, $rz(\theta_z)$. On the IBMQ experiments the angles were varied from $\theta_y \in [0, \pi]$ and $\theta_z \in [0, 2\pi]$. For both $\theta_y$ and $\theta_z$ we create $17$ linearly spaced (inclusive) angles to vary the message qubit state over, resulting in $289$ message qubit states.\footnote{Note that linearly spaced angles are not a geometrically accurate representation of the Bloch sphere, cf.~\cite{single-qubit-state-tomography}, but sufficient for characterizing the algorithm and hardware performance of our quantum telecloning circuits.} Then each of these circuits are run $3$ times for each of the three Pauli basis states for the state tomography estimation \cite{Qiskit, PhysRevLett.108.070502}; therefore the total number of telecloning circuits is $867$. Additionally, we run measurement error mitigation circuits in the same job as the quantum telecloning circuits. As an example, for a deferred measurement PCC job submission the final number of circuits combined into a single job is $867 + 2^2 = 871$. However, for a post selection PCCC job submission, the total number of circuits combined into a single job is $867 + 2^5 = 899$. The ibmq\_montreal and ibmq\_toronto backends are currently the two IBMQ devices which allow the largest number of circuits to be submitted in a single job across all IBMQ devices; that limit is $900$ circuits per job. Each post selection state (which constitutes a different circuit for each state) is run separately, but with its own measurement error mitigation circuits. Therefore, this methodology fits within the provided system constraints. It is important to run all circuits within the same time span in order to attempt to mitigate the effects of noise drift and to get consistent results. All IBMQ circuits used $30,000$ shots per circuit in order to get statistically robust sample sizes. 

For the post selection implementation, in order to compute the fidelity of each clone qubit, the results are first filtered for the post-selected states, and then the counts of 0 and 1 measurements for each clone qubit (measured using parallel state tomography) are summed across the 4 post-selection states. Then a single fidelity value is computed using the sum of the post selected states. Interestingly, the proportion of post-selection results which are kept for each bitstring varies between the PCC and APCC circuit implementations; the APCC circuit implementation have an ideal proportion that is kept of $0.25$ for each of the $4$ classical results. Whereas the PCC circuit implementation has post selected samples proportion that varies between $\frac{1}{3}$ and $\frac{1}{6}$. This is further examined in the experimental results in Section \ref{sec:results-post-selection-proportions}. 

Circuits executed on the Quantinuum H1-2 backend $11$ linearly spaced $\theta_y$ angles over $[0, \pi]$ were used, along with a fixed $rz(\frac{\pi}{2})$ gate. Each circuit, using parallel state tomography, was run using $300$ shots. The small number of shots is due to usage limitations in place on the Quantinuum system. By default, the Quantinuum backend applies some circuit optimizations in the process of compiling the user-submitted circuit to the device hardware; this flag is left on when executing circuits.

\section{Results}
\label{sec:results}
In this section we describe the measured fidelities of the different quantum telecloning implementation variants run on both IBMQ and Quantinuum NISQ computers. The fidelity results can vary significantly, and in order to appropriately show the relative fidelity values achieved, we have scaled the figures into the same fidelity ranges (in terms of either fidelity axis or heatmap values) among all two clone figures and separately among all three clone figures. The differentiation between clone numbers is due to the decrease in theoretical clone fidelity with increasing $M$ (see Eq.~\ref{eq:theoretical-fidelity}).


\begin{figure*}[t!]
    \centering
    \includegraphics[width=0.46\textwidth]{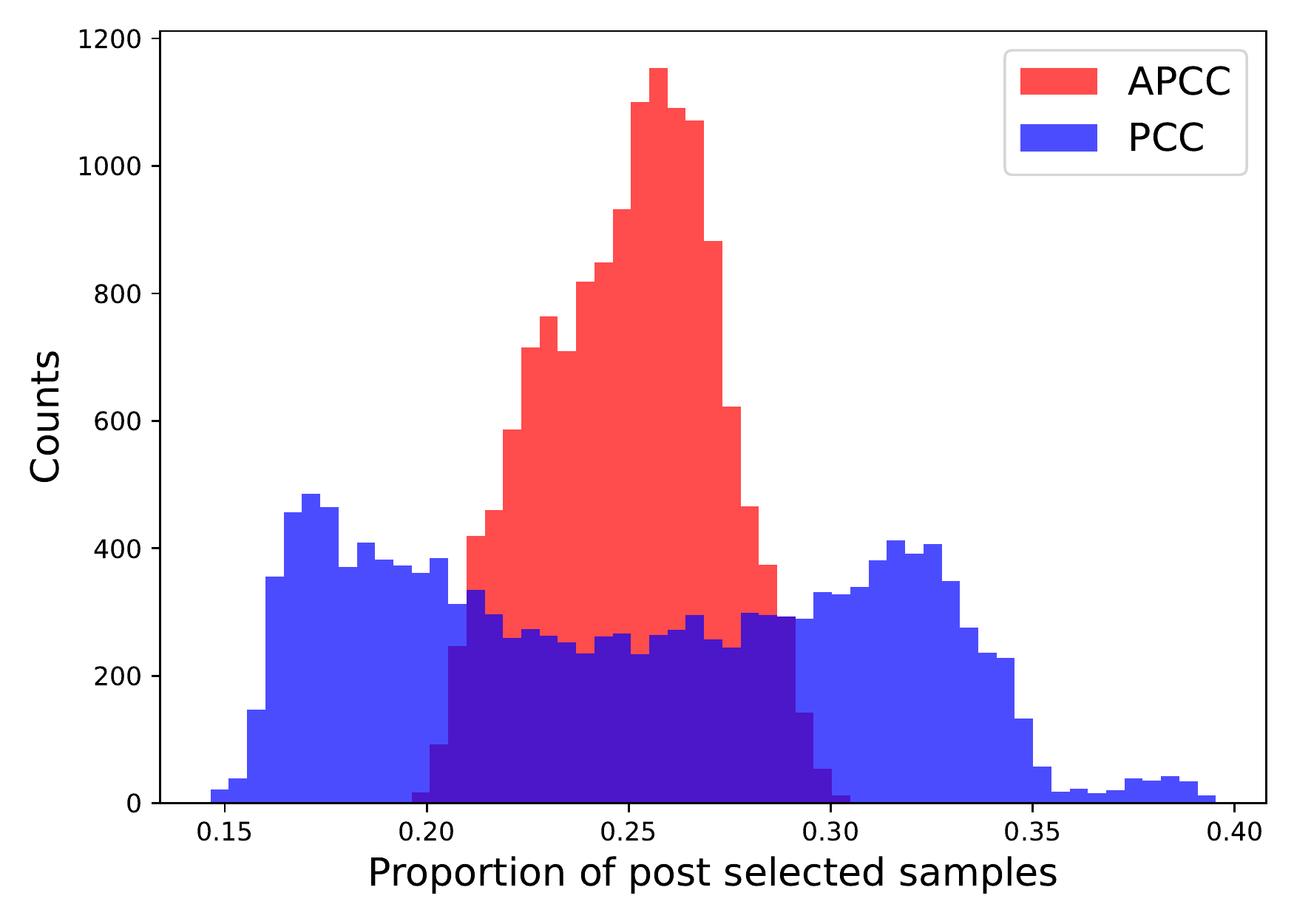}
    \includegraphics[width=0.46\textwidth]{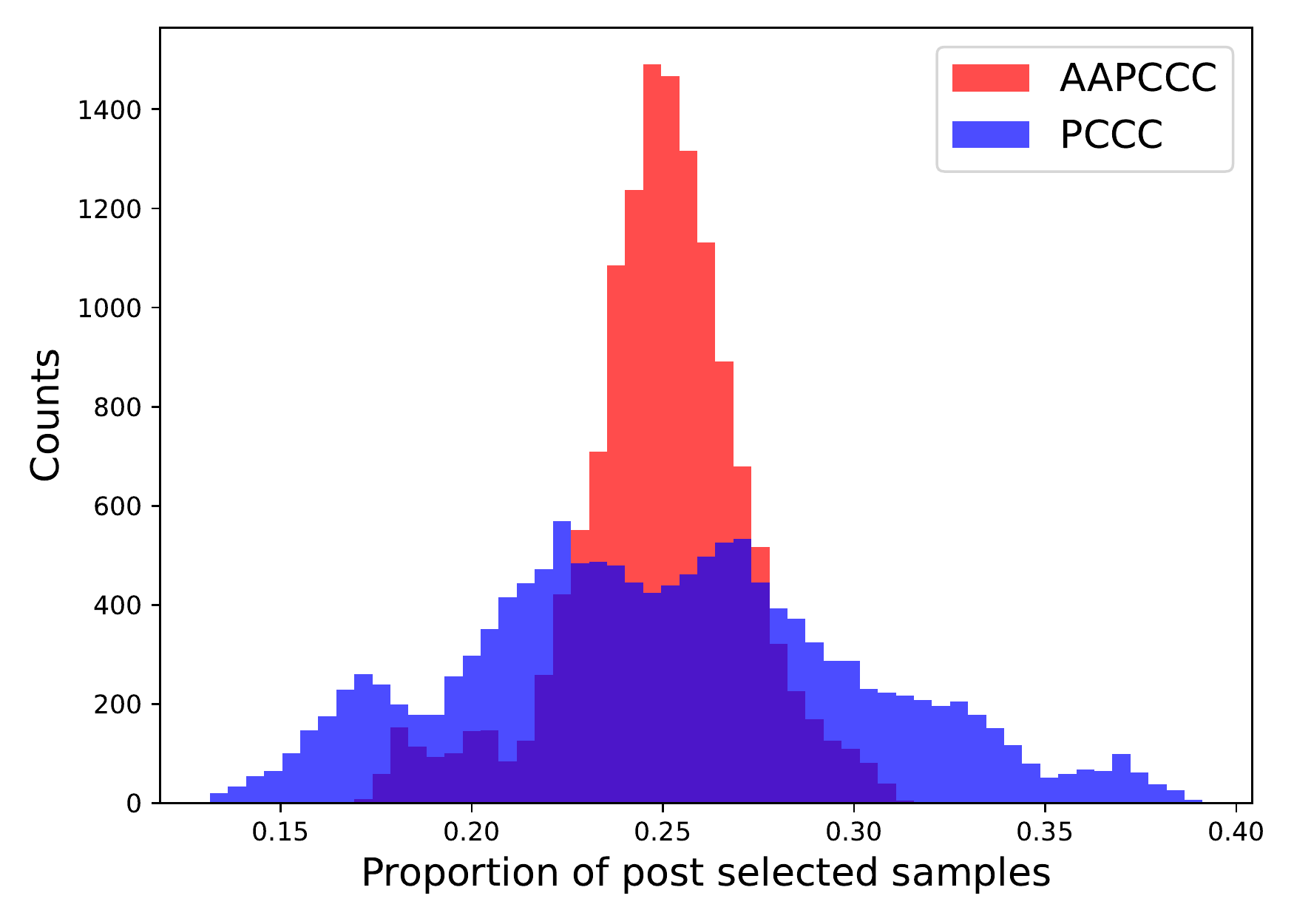}
    \caption{Post selection count histograms for two clone circuits (left hand) and three clone circuits (right hand) across $4$ separate runs of each circuit on different qubit layouts of ibmq\_montreal. These are the raw circuit measurement counts, not corrected by the measurement error mitigation procedure. }
    \label{fig:post_selection_counts_histogram}
\end{figure*}

Figure \ref{fig:PCC} shows the measured fidelity values in the form of a heatmap for the deferred measurement and post selection variants of the PCC circuit with and without measurement error mitigation. Figure \ref{fig:APCC} shows the fidelity results for the APCC circuit, Figure \ref{fig:PCCC} shows the fidelity results for the PCCC circuit, and Figure \ref{fig:AAPCCC} shows the fidelity results for the AAPCCC circuit. In order to visually differentiate the two clone heatmaps (Figures \ref{fig:APCC} and \ref{fig:PCC}) from the three clone heatmaps (Figures \ref{fig:AAPCCC} and \ref{fig:PCCC}), two different color gradients are used. Across all four of these figures there are several consistent findings in the data. First, measurement error mitigation consistently results in higher mean fidelities compared to no measurement error mitigation. Second, post selection results in higher clone fidelity compared to deferred measurement with the exception of the AAPCCC circuit where the mean deferred measurement fidelity was slightly better than the mean post selection fidelity. Third, while the clones are theoretically symmetric, the heatmap fidelities show some noticeable differences in the fidelity trends across different message qubit states. Fourth, increasing circuit depth and number of qubits used in the circuit corresponds to a decrease in the measured fidelity - meaning that the PCC circuit had the best mean fidelity compared to the theoretical maximum, whereas the AAPCCC circuit with the largest gate depth had the worst mean fidelity compared to the theoretical maximum. 

Measurement error mitigation reduces the effective error rate for noisy measurements therefore making the simulations more accurate. Deferred measurement utilizes more CNOT gates compared to post selection thus incurring more noise during circuit execution. And larger circuit sizes use more gates, have higher circuit depth, and use more qubits thus also incurring more noise during circuit execution. Thus, each of these four observations of the data are consistent with what we expect from executing these circuits, with the exception of the differences in clone fidelity as this was not an expected result given that the theoretical telecloning circuit is symmetric.

Following from the observed fidelity pattern differences between clones, Figures \ref{fig:PCC}, \ref{fig:APCC}, \ref{fig:PCCC} and \ref{fig:AAPCCC} all show clear clusters and patterns of fidelity (or infidelity). In particular, this seems to show that the clone fidelity of the universal quantum telecloning circuits are state \emph{dependent} when implemented on the IBMQ hardware (however, the logical circuits themselves are universal, meaning state \emph{independent}). The exact cause of this mechanism is not known, but it could be because of asymmetry in the quantum control mechanisms or persistent environmental noise. Similar trends in single qubit fidelity have been observed before on IBMQ hardware \cite{single-qubit-state-tomography} and the telecloning process could be cloning the already slightly decoherent message qubit, which may decohere or be prepared in an unintentionally state dependent manner.

Figure \ref{fig:Quantinuum_results} shows the fidelity (y-axis) results for varying $ry$ angles (x-axis) in the range $[0, \pi]$ for the PCC, APCC, PCCC, and AAPCCC circuits implemented on the Quantinuum H1-2 device. The fidelity results are consistently approaching the theoretical clone fidelity values. These high fidelity results with no error mitigation can be attributed to the low error rates \cite{pino2021demonstration}, all-to-all connectivity, and the capability of mid-circuit measurement with if statements on the H1-2 device. 

\begin{figure*}[t!]
    \centering
    \includegraphics[width=0.46\textwidth]{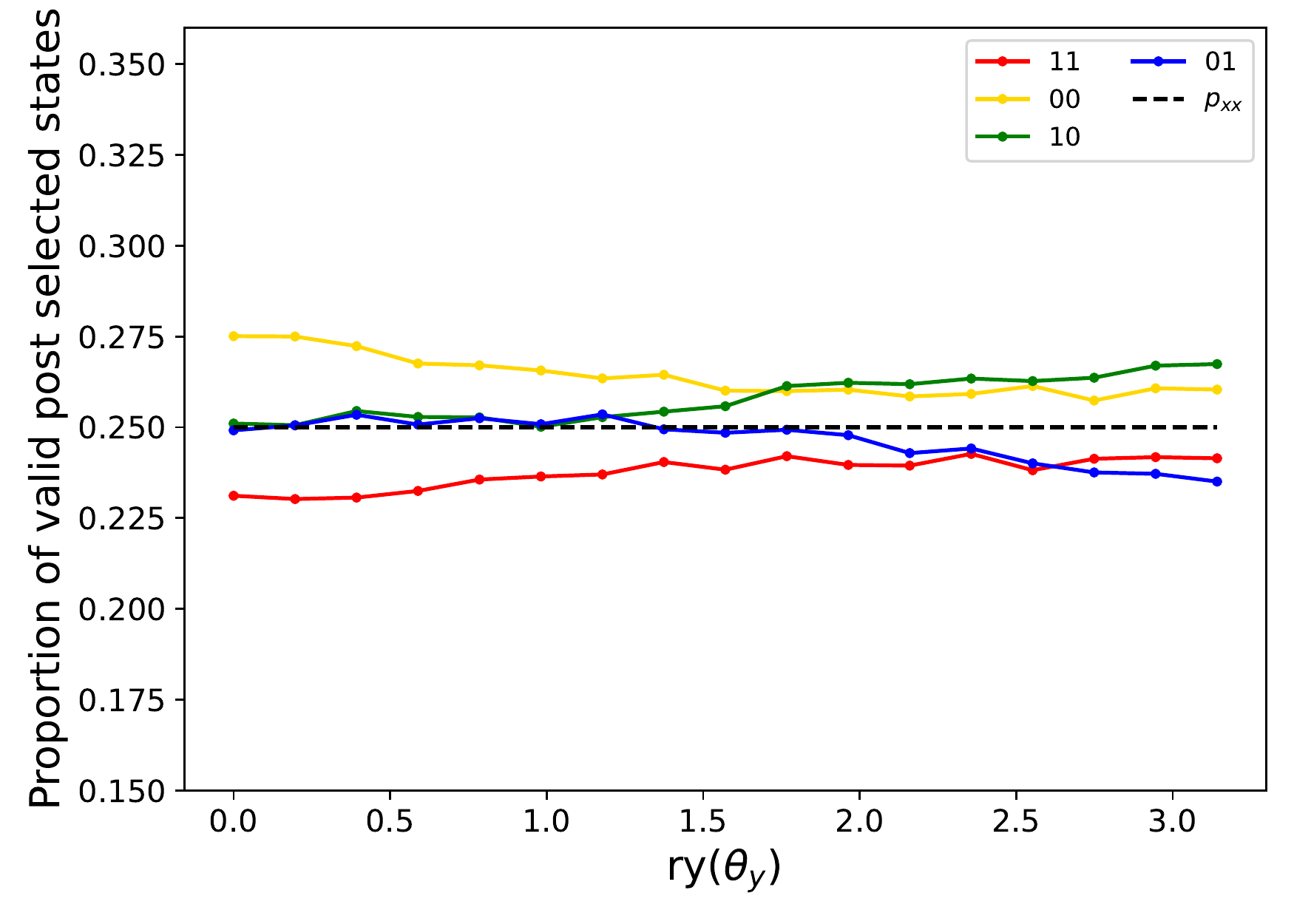}
    \includegraphics[width=0.46\textwidth]{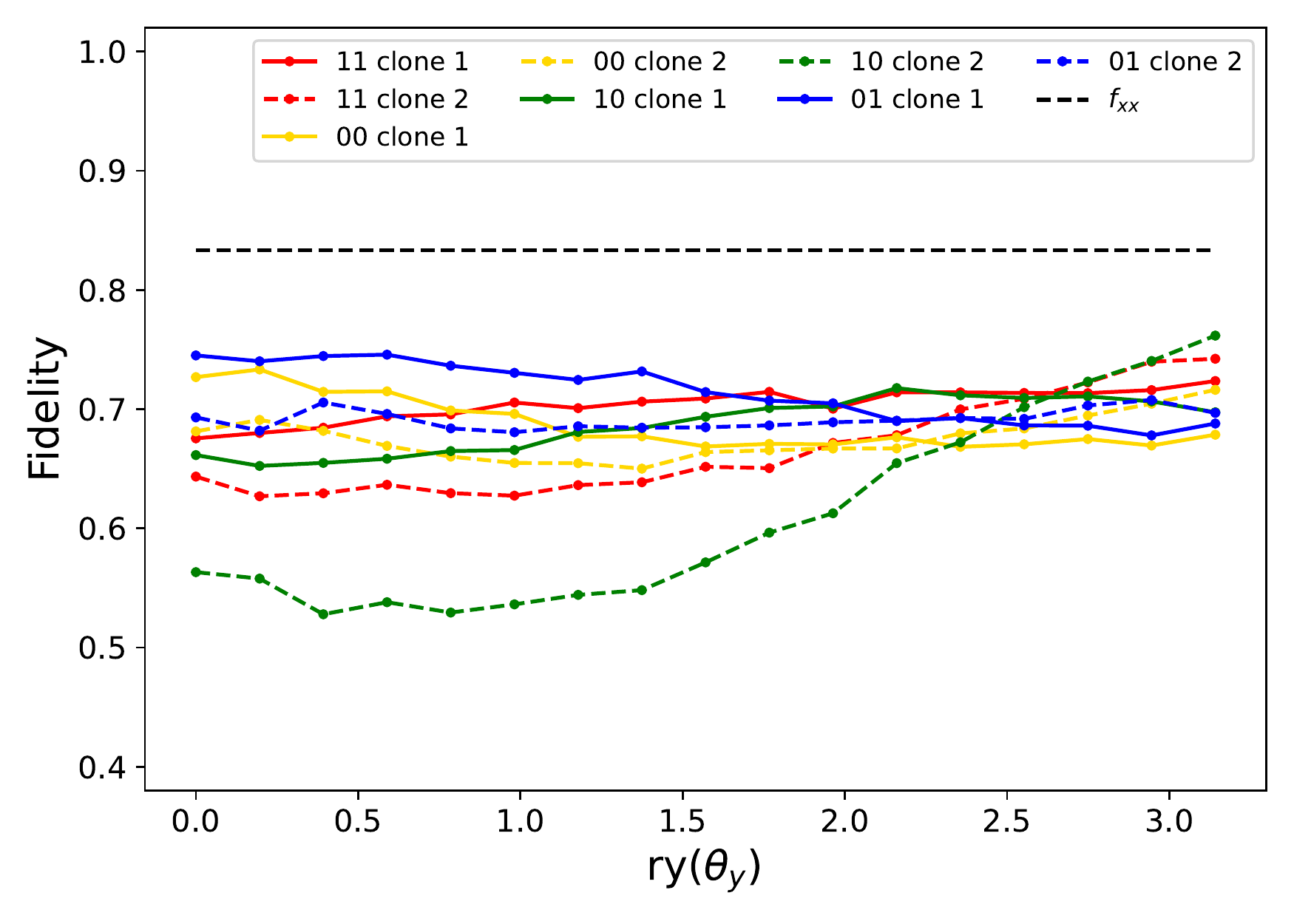}
    \includegraphics[width=0.46\textwidth]{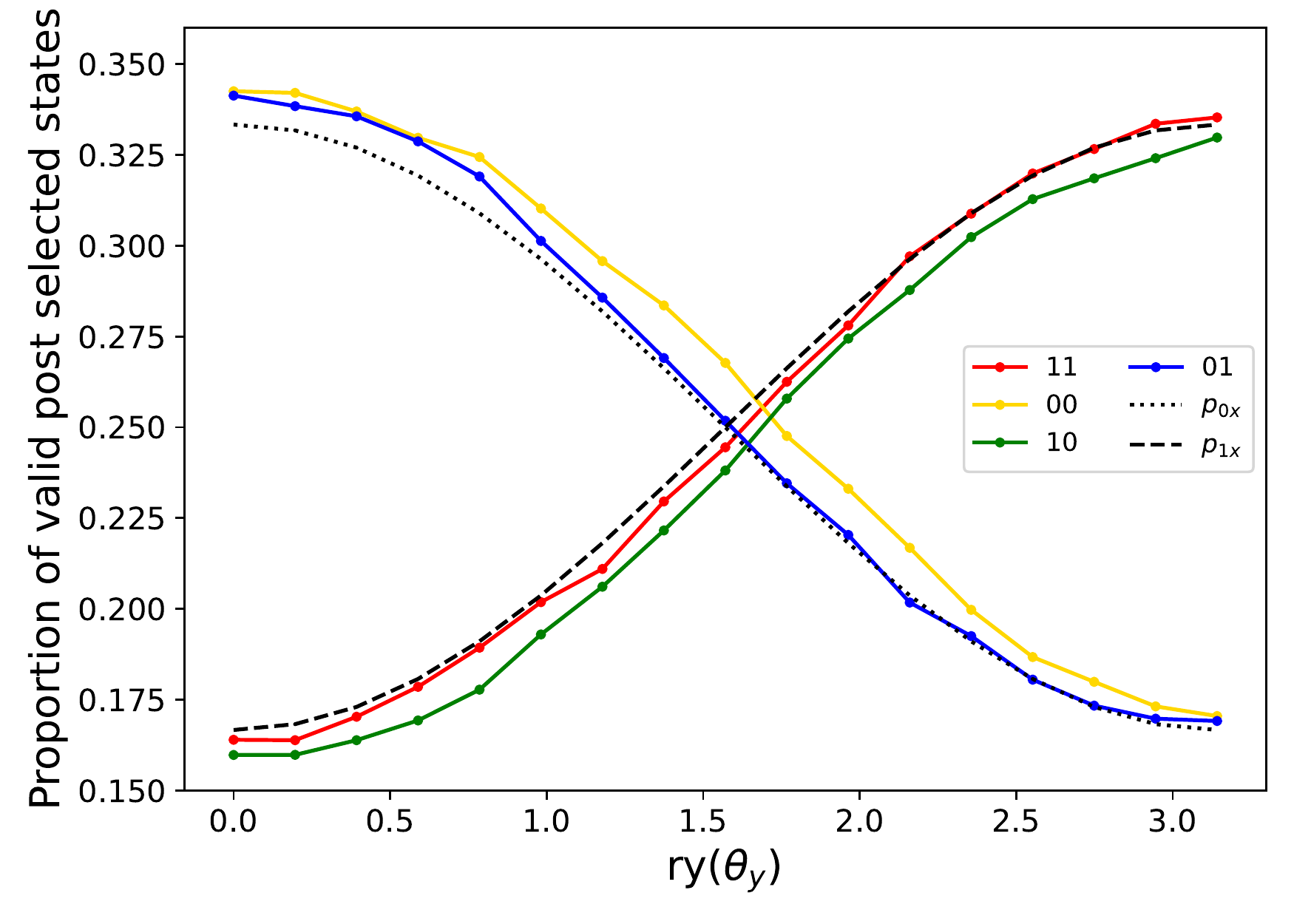}
    \includegraphics[width=0.46\textwidth]{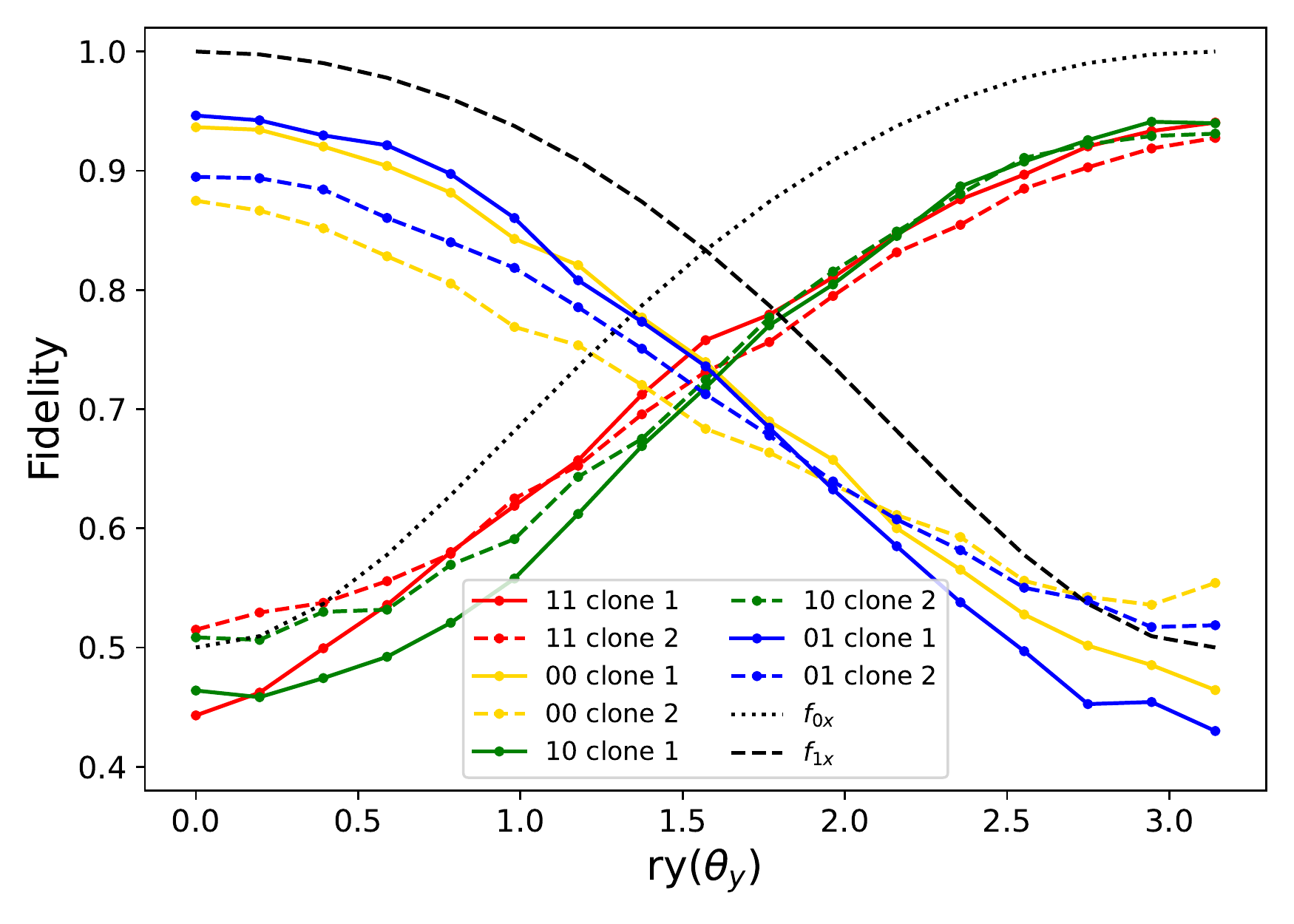}
    \caption{A more detailed analysis of post selection results in terms of samples kept after post selection and fidelity for each of those post selected state samples (no measurement error mitigation). The x-axis is varying $\theta_y \in [0, \pi]$, and $\theta_z = \frac{\pi}{2}$. Top row shows results from the APCC circuit, and the bottom row shows results from the PCC circuit (both executed on ibmq\_montreal). Left hand column shows the proportion of samples which were kept after post selection. Right hand column shows the fidelity results for the post selection experiments split by the $4$ post selection states (as well as by clone number). These results come from a subset of the post selection experiments shown in Figures \ref{fig:PCC} and \ref{fig:APCC}.}
    \label{fig:post_selection_counts_varying_ry}
\end{figure*}

\subsection{Post selection characteristics}
\label{sec:results-post-selection-proportions}
Figure \ref{fig:post_selection_counts_histogram} shows the distribution of post selection states for PCC, APCC, PCCC, and AAPCCC circuits. The total number of data points being plotted in each distribution is $13872$.\footnote{The number of data points comes from $3$ parallel state tomography circuits, $17$ $\theta_y$ and $\theta_z$ angles each, $4$ post selection states, and $4$ different hardware implementations: $3 \cdot 17 \cdot 17 \cdot 4 \cdot 4 = 13872$}
The APCC distribution is expected to be approximately binomial with a mean at $0.25$. The wide variance observed is due to the overall noise from implementation on the NISQ hardware. 

Figure \ref{fig:post_selection_counts_varying_ry} shows the post selection state proportion and fidelity for each post selection state as a function of $ry(\theta_y)$. Notably the proportion of samples which are kept after post selection are relatively consistent with the (expected) theoretical values. The two possible ideal post selection proportions $p_{00}=p_{01}$ and $p_{10}=p_{11}$ for the PCC circuit are given in Equation \eqref{eq:PCC_post_selected_proportion}, and the two corresponding ideal post selected fidelities $f_{00}=f_{01}$ and $f_{10}=f_{11}$ are given by Equation \eqref{eq:PCC_post_selected_fidelity}:



\begin{align} 
p_{0x} = \frac{2  \cos(\tfrac{\theta}{2})^2 + \sin(\tfrac{\theta}{2})^2}{6},
&\ p_{1x} = \frac{\cos(\tfrac{\theta}{2})^2 + 2 \sin(\tfrac{\theta}{2})^2}{6} 
\label{eq:PCC_post_selected_proportion} \\
f_{0x} = \frac{4 \cos(\frac{\theta}{2})^2 + \sin(\frac{\theta}{2})^2}{12\ p_{0x}},
&\ f_{1x} = \frac{\cos(\frac{\theta}{2})^2 + 4 \sin(\frac{\theta}{2})^2}{12\ p_{1x}}
\label{eq:PCC_post_selected_fidelity}
\end{align}

The theoretical curves and the experimental curves for PCC intersect at the same post selection proportion, and the fidelity results also intersect at the expected angle of $\frac{\pi}{2}$ although at a lower fidelity value than the theoretical curves (which is consistent with the fidelity results in Figures \ref{fig:APCC} and \ref{fig:PCC}).

Figures \ref{fig:post_selection_counts_histogram} and \ref{fig:post_selection_counts_varying_ry} show that the proportion of post selected states for PCC varies between $\frac{1}{6}$ and $\frac{1}{3}$, while for APCC they are a constant and equal-weighted $\frac{1}{4}$. The measured fidelity for each post selection experiment is effectively a weighted sum (weighted by the proportion of post selected states) of the fidelities for each post selection state.

\section{Conclusion}
\label{sec:conclusion}
In this work we demonstrated the viability of performing an emulation of quantum telecloning on NISQ computers. The most direct method of performing quantum telecloning on a gate model quantum computer is to perform the necessary bell measurement during circuit execution, and then use real time classical conditional control in order to generate the clones in real time. We demonstrate that this is possible on the Quantinuum H1-2 device. In cases where this feature is not present, we show that alternative methods such as deferred measurement and post selection allow for analysis of the telecloning protocol on ibmq\_montreal. We also present a novel $1 \rightarrow 3$ telecloning quantum circuit with no ancilla. Importantly, with the use of measurement error mitigation in the case of deferred measurement and post selection on the IBMQ devices, the clone fidelities can approach the theoretical fidelity limit. The low error rate and all-to-all connectivity of the Quantinuum H1-2 device cause correspondingly high fidelity values which are consistently close to the theoretical fidelity limit. 
There are numerous research avenues left for future work:

\begin{itemize}
    \item How do circuits with telecloning states with ancilla for $M \geq 4$ clones perform on current NISQ devices? 
    \item Does a telecloning state for $M \geq 4$ without ancilla exist?
    \item Given the availability of cloud based NISQ platforms, is there meaningful quantum cryptographic analysis \cite{https://doi.org/10.48550/arxiv.2012.11424, Bartkiewicz_2013, PhysRevA.88.012331} or experimental quantum network protocol implementations that can be performed on these devices?
    \item Circuits of the related idea of remote information concentration protocols could also be implemented on NISQ computers \cite{PhysRevLett.86.352, wang2013manytoone}. 
\end{itemize}

\section{Acknowledgments}
\label{sec:acknowledgments}

The authors thank Adrien Suau for helpful discussions on single qubit tomography. 
We acknowledge the use of IBM Quantum services for this work. The views expressed are those of the authors, and do not reflect the official policy or position of IBM or the IBM Quantum team.
This research used resources of the Oak Ridge Leadership Computing Facility, which is a DOE Office of Science User Facility supported under Contract DE-AC05-00OR22725.

\bibliographystyle{plainurl}
\bibliography{references.bib}{}

\end{document}